\documentclass[journal=jacsat,manuscript=article]{achemso}

\usepackage[version=3]{mhchem} 

\author{Harshad Bhapkar}
\author{Pierre Kawak}
\email{pskawak@gmail.com}
\author{David S. Simmons}
\email{dssimmons@usf.edu}
\affiliation[USF]
{Department of Chemical, Biological, and Materials Engineering, The University of South Florida, Tampa, FL 33612}

\title[An \textsf{achemso} demo]
  {Anisotropic Nanoparticle Rejamming Triggers Thermodynamic Cavitation in Elastomer Nanocomposites}

\abbreviations{IR,NMR,UV}

\begin{document}


\begin{abstract}

Nanoparticles can dramatically reinforce elastomers while paradoxically causing cavitation at lower strains. Despite decades of research, the microscopic origin of this behavior has remained unsettled. Here, molecular dynamics simulations reveal that cavitation and failure arise from the nanoparticulate reinforcement mechanism itself. Initial jamming of the nanoparticulate network leads to a buildup of negative pressure in the elastomer matrix,  reinforcing it and simultaneously driving it towards a cavitation limit. This crisis is initially averted by yield of the particle network. However, an anisotropic rejamming event of the nanoparticles ultimately drives a runaway negative pressure buildup that leads to cavitation and failure. These results identify nanoparticle-network-induced thermodynamic cavitation as the origin of void formation in elastomeric nanocomposites, and they establish collective filler dynamics as a potential point of control of ultimate failure. 
  
\end{abstract}

\section{Main Text}

Introducing nanoparticles to elastomeric nanocomposites can yield massive reinforcement,\cite{kumar_nanoparticle_2017,Kumar2017a, song_concepts_2016, hamed_reinforcement_2000} yet paradoxically often leads to cavitation and failure at lower strains.\cite{peddini_nanocomposites_2015,gent_failure_1984,Dannenberg1986}  For over 50 years, this has most commonly been understood via a picture of defect-driven failure -- in part because it was not clear how nanoparticles could drive microscopic cavitation in the absence of preexisting defects.\cite{gent_internal_1957,gent_cavitation_1990,ball_discontinuous_1982,gent_failure_1984,gent_fracture_1991,williams_spherical_1965,eshelby_determination_1957,lake_strength_1967, pardoen_extended_2000, gologanu_theoretical_2001, griffith_vi_1921, gent_surface_1969, cho_cavitation_1988} This classic scenario of void formation in elastomeric nanocomposites posits that either particles induce local stress concentrations that drive growth of pre-existing defects or that particle surfaces serve as nucleating sites for void formation. This picture, however, has left multiple major questions unresolved. First, efforts to observe the pervasive microscopic defect structures postulated by these theories have not generally succeeded\cite{creton_fracture_2016,gent_fracture_1991, gent_new_1996, gehant_criteria_2003, marr_void_1997, heinrich_reinforcement_2002, beutier_situ_2022, oakey_influence_1999} while spontaneous emergence of nanovoids apart from any preexisting defects has also been observed.\cite{zhang_nanocavitation_2012} Second, it is difficult to understand how nanoparticle surfaces could seed voids in systems with strong particle-polymer attractions that preclude interfacial debonding.\cite{chang_creep_2021,poulain_damage_2017,ilseng_experimental_2017,oberth_tear_1965,sorkin_atomistic-scale_2021} Third, it remains unclear why the same nanoparticles that favor earlier cavitation nevertheless reinforce the elastomer. A complete microscopic theory of nanoparticle-induced reinforcement and failure must explain not only why voids form, but why the same nanoparticles that induce reinforcement also appear to favor cavitation and failure.

Here, we report on molecular dynamics simulations indicating that void formation in these systems instead occurs via thermodynamic cavitation, caused when the reinforcement mechanism drives the polymer matrix beyond its intrinsic thermodynamic cavitation limit. This event is triggered by the onset of an anisotropic nanoparticulate rejamming event that causes a runaway negative pressure buildup in the elastomer matrix phase. These results fundamentally recast the nature of void formation and failure in elastomeric nanocomposites while suggesting the possibility of failure-condition control through modulation of high-strain nonaffine nanoparticle motion.

To probe the mechanisms of cavitation in elastomeric nanocomposites, we perform molecular dynamics simulations\cite{plimpton_fast_1995,kawak_amdat_2026,simmons_amorphous_2025,stukowski_visualization_2010} of sub-entangled crosslinked elastomers reinforced with nanoparticle clusters constructed from sintered icosahedra, following a protocol developed in our prior works.\cite{smith_horizons_2017,smith_poisson_2019,kawak_central_2024} We probe systems over a range of cluster structure, ranging from 7 particles per sintered cluster (denoted NP7) to 13 particles per sintered cluster (denoted NP13) (Figure S1). Simulations are performed at a high temperature and with thermodynamically neutral polymer-particle interactions, such that glassy polymer layers that can be induced by strong polymer-particle interactions at low temperatures are not present. 

\begin{figure*}[!htp]
  \centering
  \includegraphics[width=\textwidth]{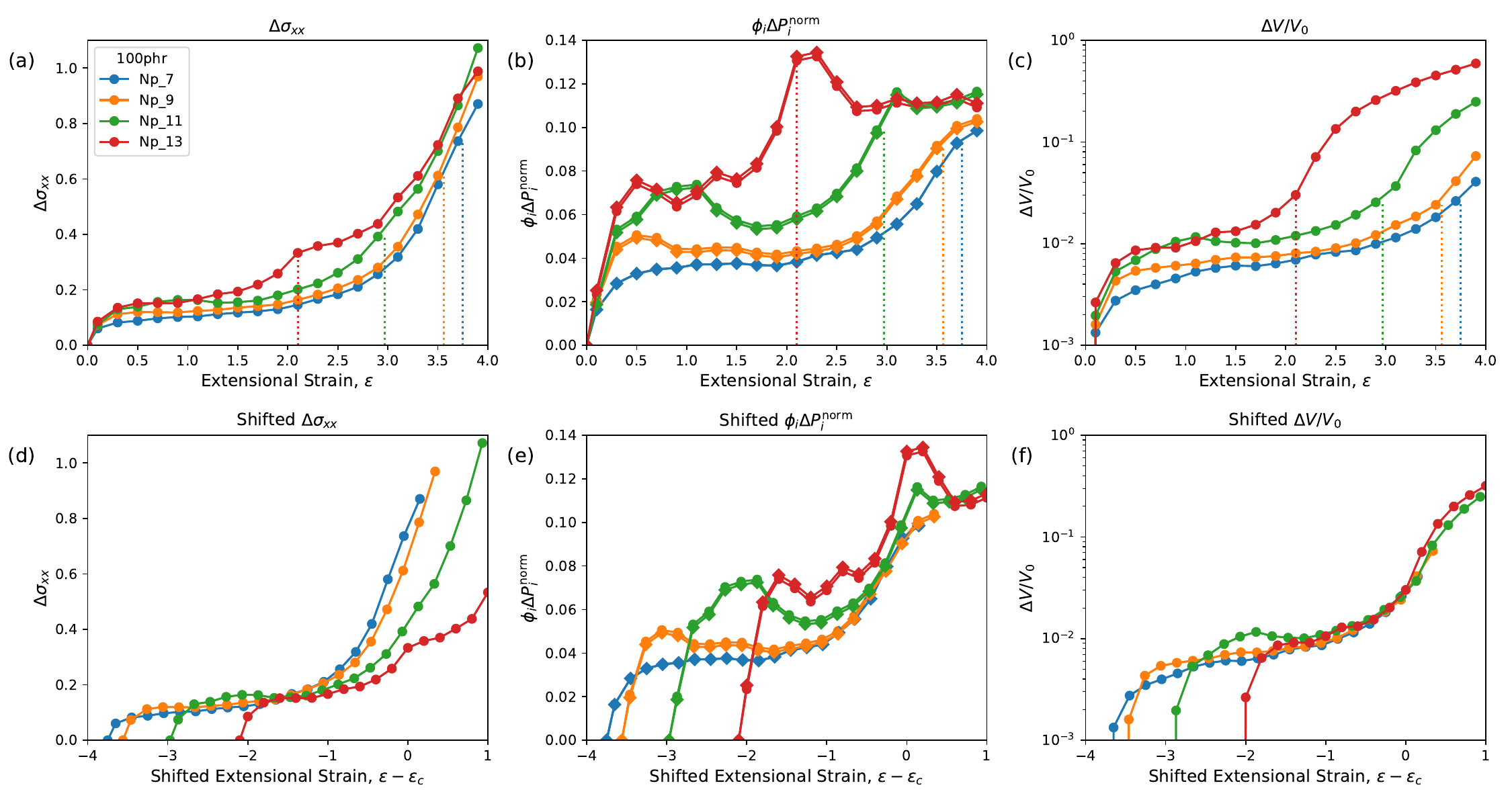}
  \caption{a) Engineering extensional stress-strain curves from extensional deformation of filled elastomers relative to the quiescent system at constant filler loading of 0.252 volume fraction (100 parts per hundred rubber) with varying filler structures consisting of 7, 9, 11 and 13 icosahedral clusters. b) Product of pressure in normal direction, $P^{\mathrm{norm}}$, and volume fraction, $\phi$ of the polymer. The negative of the polymer pressure is plotted and it super imposes on the positive filler pressure. c) Change in the volume of the system relative to the pre-stretching volume normalized by the initial volume showing the onset of void formation. The dotted lines indicate the corresponding cavitation strain, $\varepsilon_c$, identified by where sustained voids are formed in the polymer matrix. d), e), and f) Shifted extensional stress, normal pressure and change in volume of the system vs shifted strain of 3.75, 3.65, 3.10, and 2.18 for NP7, NP9, NP11, and NP13 systems respectively.
  }
  \label{fig:stress_strain_nps_100phr}
\end{figure*}

The low-strain portion of the results in Fig.~\ref{fig:stress_strain_nps_100phr}a-c establishes a key baseline for a new understanding of cavitation in elastomeric nanocomposites. As seen here, enhancement of the low-strain stress response with increasing particle structure is accompanied by increasing positive volume deviations (reflecting a suppression of Poisson's ratio for the composite, Fig. S2a) and by emergence of an internal normal pressure balance between polymer (under negative pressure) and particle network (under positive pressure). This pattern of behavior is consistent with the recent finding that low-strain reinforcement in elastomeric nanocomposites is fundamentally driven by a volumetric tug of war between particle network and polymer network -- the former prefers to grow in volume under deformation, while the latter prefers to conserve volume.\cite{smith_poisson_2019,kawak_origin_2025,kawak_glassy_2026} The resulting intermediate compromise involves a Poisson's ratio suppressed below 0.5 and corresponding growth in composite volume under deformation. This volume growth gives rise to a large negative isotropic pressure within the elastomer matrix (driven by the elastomer's bulk modulus), which is the origin of low-strain reinforcement.

The emergence of a negative polymer partial pressure within the polymer matrix suggests a potential origin of cavitation in elastomeric nanocomposites under strain. If these internal stresses continue to rise due to ongoing competition between filler and polymer volumes, the polymer's negative pressure must eventually exceed its thermodynamic cavitation pressure. If so, this would draw a direct connection between the mechanism that drives reinforcement and the mechanism that drives early failure in elastomeric nanocomposites. However, as seen in Fig.~\ref{fig:stress_strain_nps_100phr}a, this crisis is initially averted. At strains in tens of percent, the system exhibits a soft yield behavior. This yield, which is known as the Payne effect,\cite{payne_dynamic_1962,harwood_stress_1965} is established to be driven by a yield of the nanoparticle network.\cite{wang_strain-induced_2005,robertson_spectral_2006,richter_jamming_2010}  Fig.~\ref{fig:stress_strain_nps_100phr}b-c reveals that this event also leads to a transient plateau in volume deviations and normal pressure, averting (at least at first) a cavitation crisis.

\begin{figure*}[!htp]
  \centering
  \includegraphics[width=\textwidth]{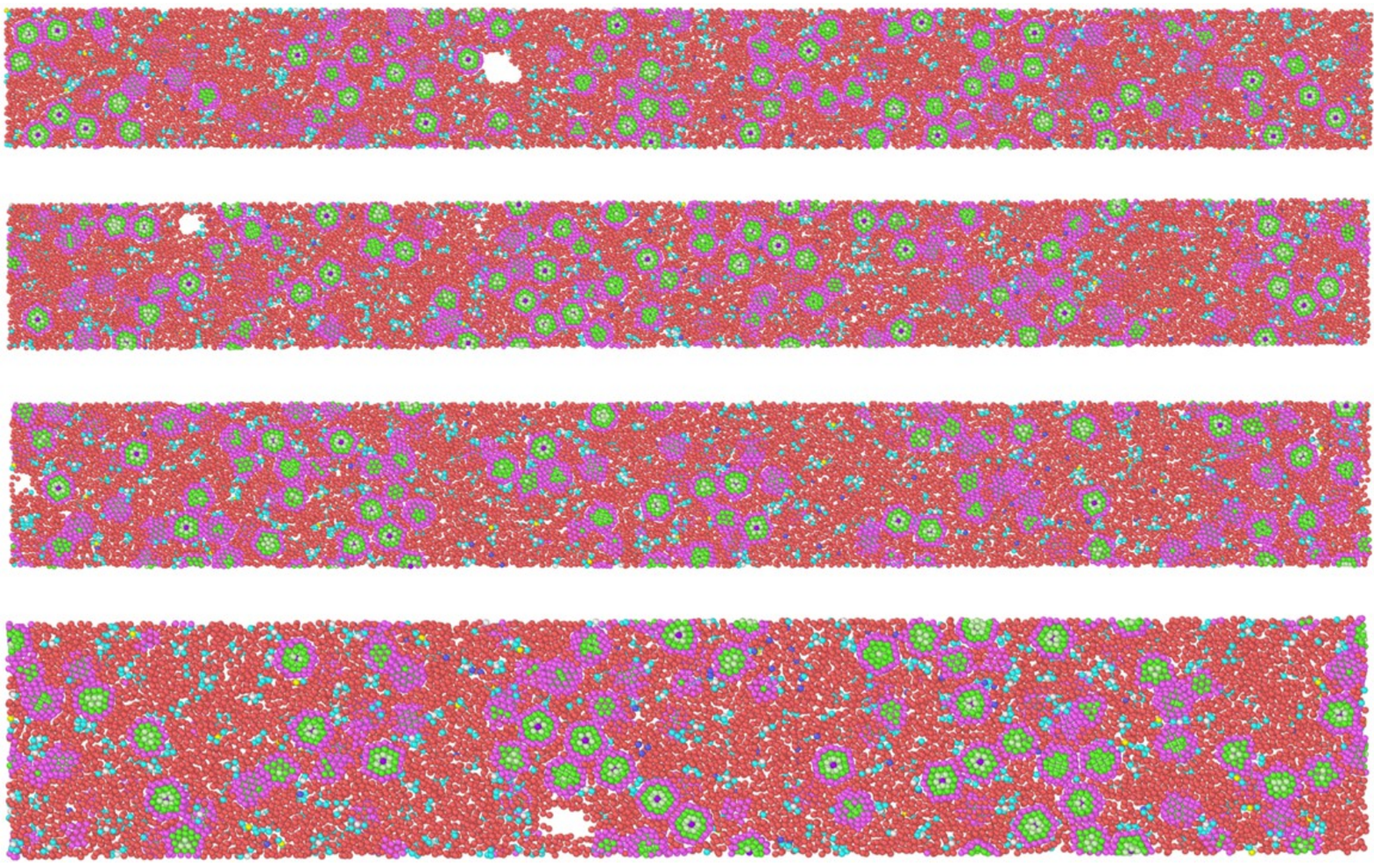}
  \caption{ Shows voids formed in the 100 PHR systems at strains following which the volume of the systems drastically increases. Using Ovito, the system was scanned manually with 3$\sigma$ slices normal to the direction of elongation. Voids were formed at strains of 3.75, 3.65, 3.10, and 2.18 for NP7, NP9, NP11, and NP13 systems indicated by panels from top to bottom, respectively.}
  \label{fig:nano_voids}
\end{figure*}

Ultimately, however, we observe that this plateau ends with a rapid upturn in elongational stress, normal stress, and volume, seen in Fig.~\ref{fig:stress_strain_nps_100phr}  at high strain. In all simulated systems, this leads rapidly to cavitation, such that further elongation of the system leads to void growth. This is immediately evident from the magnitude of volume change -- no condensed liquid (or soft solid) far from its vapor-liquid critical point can sustain volume increases on the order of tens of percent without cavitation.  Indeed, noting that rapid volume growth begins in all systems by the time volume deviations have reached 3\%, we observe in Fig.~\ref{fig:nano_voids}  that each system indeed possesses one or more small visible voids by shortly after this condition. This raises two questions. First, precisely what mechanism ultimately causes cavitation when this pressure balance and volume upturn occurs? Second, what triggers this partial pressure upturn?

To begin answering the first of these questions, we employ the approximate 3\% volume deviation criterion reported above, and we shift the x-axis of all the datasets in Fig.~\ref{fig:stress_strain_nps_100phr}a-c by the strain $\varepsilon_c$ at which this volume deviation is reached. As can be seen in Fig.~\ref{fig:stress_strain_nps_100phr}d-f, the volumetric strain data Fig.~\ref{fig:stress_strain_nps_100phr}f (and Poisson ratios, Figure S2b) for all systems form master curves over a large range of high strains spanning at least $\pm$ 100\% strain around $\varepsilon_c$, indicating that the form of the volume growth through the onset of cavitation is universal across these four systems. Moreover, the volume-normalized species-specific normal stresses of the four systems (Fig.~\ref{fig:stress_strain_nps_100phr}e) exhibit a near-collapse in the strain-hardening regime based upon this strain shift. Physically, this suggests that the four systems exhibit a nearly uniform internal polymer normal pressure at which cavitation occurs, in the vicinity of -0.1 Lennard-Jones pressure units. Remarkably, the normal pressure at which cavitation occurs corresponds closely to typical cavitation pressures in elastomers. A Lennard-Jones reduced pressure of 0.1 corresponds to approximately 4 MPa, compared to an experimental cavitation pressure of 4.3MPa for carbon-black filled SBR.\cite{hamdi_fracture_2014} This indicates that void formation in these systems is driven by thermodynamic cavitation. Moreover, given the similarity to experiment in terms of these systems' particle structure, degree of reinforcement\cite{smith_horizons_2017,smith_poisson_2019} and cavitation pressure, this suggests that void formation in elastomeric nanocomposites can be fundamentally understood as a thermodynamic cavitation event driven by the reinforcement mechanism itself.

However, the existence of a cavitation-free plateau in volume and internal normal pressure balance prior to this event raises a final question -- what triggers the final pressure and volume growth that drives cavitation and failure?  Specifically, if the intermediate-strain regime following the Payne-effect yield continued indefinitely, the polymer pressure would never exceed the apparent threshold for cavitation (or would do so at a much larger strain). Instead, the normal pressure begins to rapidly grow again at higher strains. Crucially, this growth precedes the approximate strain of cavitation. This can be seen in Fig.~\ref{fig:stress_strain_nps_100phr}(e) and (f): volume begins to grow precipitously around a shifted strain of $\varepsilon-\varepsilon_c=0$ (by construction), whereas the normal pressure balance begins to grow by a shifted strain of -0.5 in all of these systems. This suggests that some event causes this new growth of internal normal stresses, which in turn leads to cavitation. Whatever this event is, it appears to be the proximate cause of the chain of events leading to cavitation and irreversible damage to the system.

\begin{figure*}[!htp]
  \centering
  \includegraphics[width=\textwidth]{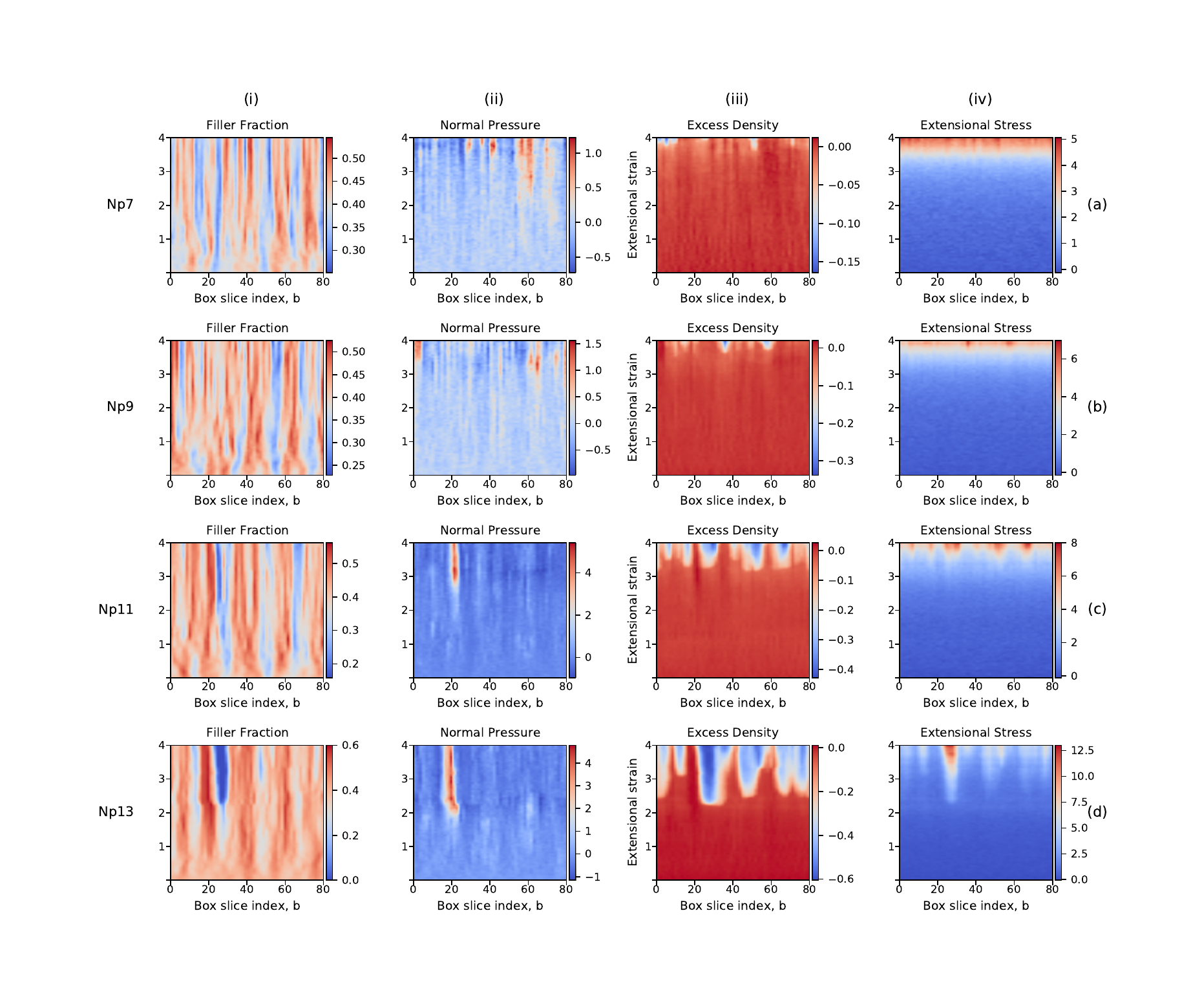}
  \caption{Plots of locally resolved properties for systems at the filler structure shown for each row. In each panel, the x-axis reports the index of the slab normal to the direction of elongation; the y-axis reports strain; and the z-axis (color) reports a local property: local filler fraction (column (i)), local normal pressure (column (ii)), local excess density relative to that expected from the polymer and volume fractions (column (iii)), and local extensional stress (column (iv)).
  }
  \label{fig:spatially-resolved}
\end{figure*}

To determine the nature of this triggering event, we spatially resolve the properties of the system into 80 equally sized rectangular slabs with normal vectors aligned along the deformation axis. As shown in Fig.~\ref{fig:spatially-resolved}~column (i), elongation of the system induces the filler particles to progressively organize into increasingly well-defined stripes aligned in the normal direction to deformation, with particle-rich regions alternating with particle-poor regions in the direction of deformation. Formation of these stripes evidently requires that the particles move in a non-affine manner under deformation.

This filler striping is followed by emergence of one or more local regions of exceptionally large positive normal pressure within the box (Fig.~\ref{fig:spatially-resolved}~column (ii)). These regions consistently correspond to a particle-enriched region, indicating that particle-rich regions increasingly bear the particle network’s normal stresses at high strain. This is followed shortly thereafter by formation of regions of large negative excess density, denoting formation of cavities (Fig.~\ref{fig:spatially-resolved}~column (iii)). Finally, at higher strains, significant elongational stress variations in the elongational direction emerge in the composite (Fig.~\ref{fig:spatially-resolved}~column (iv)), indicating an onset of mechanical instability. All of these events are most pronounced in the more highly reinforced systems comprised of more highly structured particles (Np11 and Np13), but their onset is still evident at higher strains in the more weakly reinforced systems.

\begin{figure*}[!htp]
  \centering
  \includegraphics[width=\textwidth]{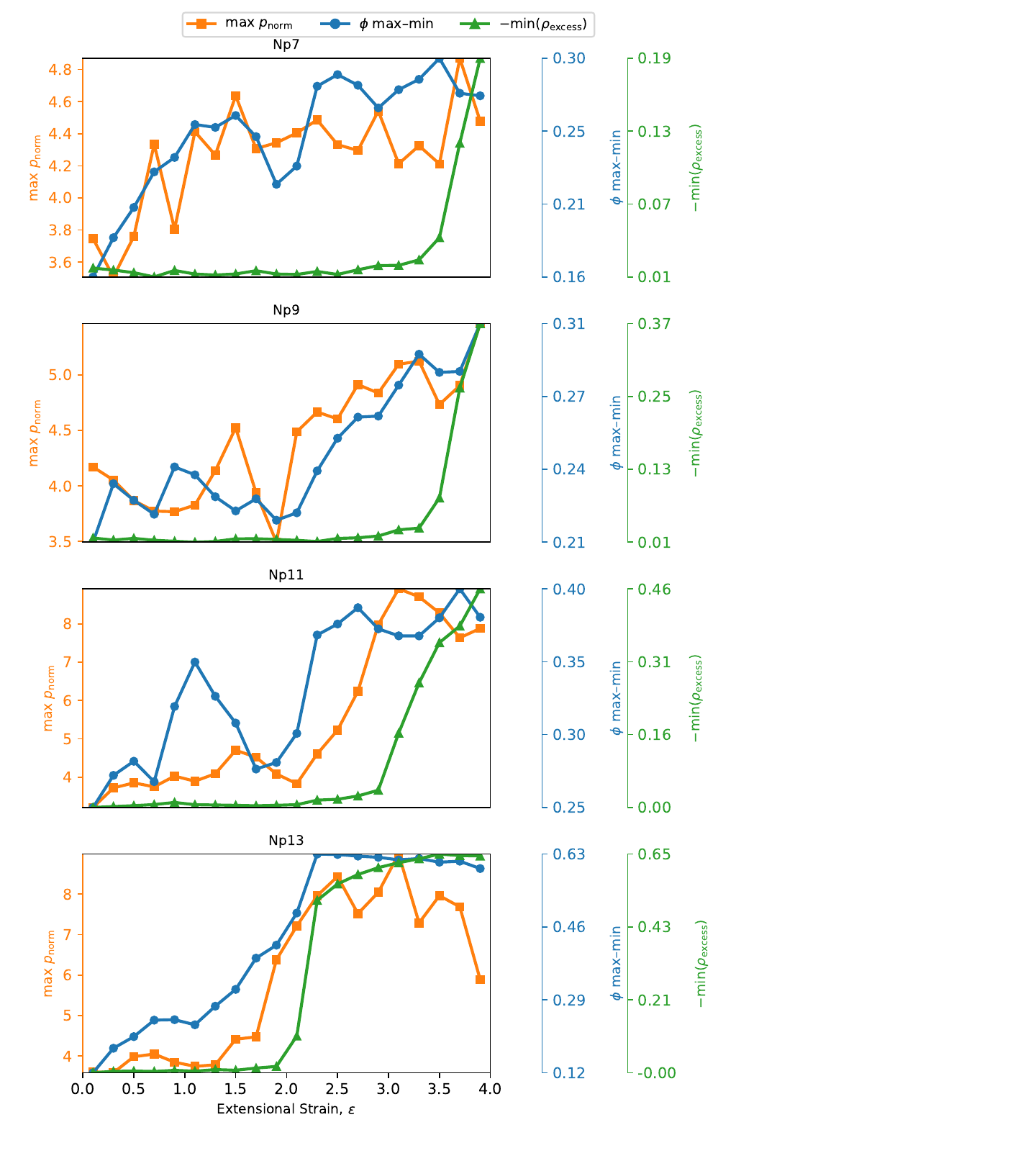}
  \caption{Extremal local behavior extracted from Fig.~\ref{fig:spatially-resolved} for systems with particles of structure NP7, NP9, NP11, and NP13, all at 100 PHR. The orange curve, max $P_{\mathrm{norm}}$ reports for each strain the largest local normal pressure taken from the spatial bins in Fig.~\ref{fig:spatially-resolved}. $\phi$ max - min reports the difference between the largest and smallest bin filler fraction at each strain in Fig.~\ref{fig:spatially-resolved}. -min$P_\mathrm{excess}$ reports the negative of minimum excess density in any bin at each strain. 
  }
  \label{fig:cascade_of_events}
\end{figure*}

These findings point to a clear cascade of events leading to cavitation and failure, highlighted by Fig.~\ref{fig:cascade_of_events}:

\begin{enumerate}
    \item Particle-enriched and particle-poor regions form during non-affine particle deformation in the intermediate strain regime (seen in Fig.~\ref{fig:cascade_of_events} via the difference in the filler fraction between the maximum and minimum of all the bins in the system).
    \item This leads to an anisotropic jamming event that causes a catastrophic internal normal pressure buildup (seen in Fig.~\ref{fig:cascade_of_events} via the maximum local normal pressure from Fig.~\ref{fig:spatially-resolved}~column (ii)).
    \item This causes cavitation when a critical negative partial polymer pressure is exceeded (seen in Fig.~\ref{fig:cascade_of_events} via the negative of the minimum excess density from Fig.~\ref{fig:spatially-resolved}~column (iii)).
    \item In all systems, mechanical instability (Fig.~\ref{fig:spatially-resolved}~column (iii)) and bond failure (Fig.~S3) only begins to emerge after this sequence of events. 
\end{enumerate}

 Crucially, SAXS studies of Schneider et al.\cite{schneider_correlation_2009,schneider_strain_2010} and Zhang et al.\cite{zhang_nanocavitation_2012} have provided direct experimental evidence for this type of anisotropic rejamming in elastomeric nanocomposites. Moreover, anisotropic jamming of post-yield granular solids is a well-established phenomenon in the granular jamming literature\cite{behringer_physics_2018,richter_jamming_2010}), but it appears to have gone unrecognized as a  trigger for cavitation and failure in elastomeric nanocomposites.

These results hearken back to early ``poker chip'' experiments, wherein thin layers of rubber were stretched at nearly zero Poisson's ratio by placing them between relatively large plates\cite{breedlove_cavitation_2024,gent_internal_1957}. Those experiments revealed extensive formation of voids, presumably favored by large isotropic negative pressures.\cite{gent_cavitation_1990} However, at the time there was no consensus mechanism by which large hydrostatic negative pressures would emerge in elastomeric nanocomposites under more typical (uniaxial or biaxial) deformation modes.\cite{breedlove_cavitation_2024} Our work identifies this mechanism and thus connects void formation during elastomer failure with the sort of hydrostatic cavitation observed in those early experiments.

These findings have profound implications for the design of highly reinforced rubber. They indicate that the same mechanism that produces reinforcement in elastomeric nanocomposites (volume growth and internal polymer-particle normal-stress competition) also predisposes the system towards lower-strain cavitation and failure. Critically, however, the low-strain reinforcement and high-strain onset of cavitation are not fully coupled. Cavitation is triggered by a high-strain anisotropic rejamming event that is gated by non-affine particle motion. Future research should explore the possibility of modulating this non-affine high strain particle motion and rejamming in an effort to combine large low-strain reinforcement with delayed-onset cavitation and failure.

\begin{acknowledgement}

This material is based upon work supported by the U.S. Department of Energy, Office of Science, Office of Basic Energy Sciences, under Award No. DE-SC0022329.

\end{acknowledgement}

\begin{suppinfo}

Simulation methods and supplementary data.

\end{suppinfo}

\bibliography{achemso-demo}

\end{document}